\documentclass[referee]{aa}
\usepackage{graphics}
\begin{document}
\title{Using Cepheids to determine the galactic abundance gradient.
III.First results for the outer disc
\thanks{Based on spectra collected at ESO - La Silla, Chile}
\thanks{Table A1 (Appendix) is only available in electronic form at the CDS
via anonymous ftp to cdsarc.u-strasbg.fr (130.79.128.5)
or via http://cdsweb.u-strasbg.fr/cgi-bin/qcat?J/A+A/(vol)/(page)}}
\titlerunning{Galactic abundance gradient}
\author{S.M. Andrievsky,
\inst{1,2}\,
V.V. Kovtyukh,
\inst{1,2}\,
R.E. Luck,
\inst{3}\,\
J.R.D. L\'epine,
\inst{4}\,\\
W.J. Maciel,
\inst{4}\,
Yu.V. Beletsky
\inst{1,2}}
\authorrunning{Andrievsky et al.}
\offprints{S.M. Andrievsky}
\institute{
Department of Astronomy, Odessa State University,
Shevchenko Park, 65014, Odessa, Ukraine\\
email:
scan@deneb.odessa.ua; val@deneb.odessa.ua; adelante@ukr.net
\and
Odessa Astronomical Observatory and Isaac Newton Institute of Chile,
Odessa Branch, Ukraine
\and
Department of Astronomy, Case Western Reserve University, 10900
Euclid Avenue, Cleveland, OH 44106-7215\\
email:
luck@fafnir.astr.cwru.edu
\and
Instituto Astron\^ {o}mico e Geof\' \i sico, Universidade de S\~{a}o
Paulo, Rua do Mat\~ao 1226, S\~ao Paulo SP, Brazil\\
email:
jacques@astro.iag.usp.br; maciel@astro.iag.usp.br}
\date{Received ; accepted }
\abstract
{As a continuation of our previous work, which concerned the radial
abundance distribution in the galactic disc over the distances 4--10 kpc
this paper presents the first results on the metallicicty in the outer
disc (R$_{\rm G} > 10$ kpc).
Based on high-resolution spectra obtained for 19 distant Cepheids
we sampled galactocentric distances from 10 to 12 kpc. Combined with
the results of our previous work on the inner and middle parts of the
galactic disc, the present data enable one to study the structure of
the radial abundance distribution over a large baseline.
In particular, we find indications of a discontinuity in the radial
abundance distribution for iron as well as a number of the other elements.
The discontinuity is seen at a galactocentric distance R$_{\rm G} = 10$ kpc.
This finding supports the results reported earlier by Twarog et al.
(\cite{twaet97}).
\keywords{Stars: abundances--stars: Cepheids--Galaxy: abundances--Galaxy:
evolution}}
\maketitle
\section{Introduction}
As shown in our previous papers (Andrievsky et al. \cite{andret02a} --
Paper~I, \cite{andret02b} -- Paper~II) the radial abundance distribution
within the region of galactocentric distances from 4 to 10 kpc is best
described by two distinct zones. One of them (inner: 4.0 kpc $<$R$_{\rm G}
< 6.5$ kpc) is characterized by a rather steep gradient, while in the
mid part of galactic disc (6.5 kpc $<$R$_{\rm G} < 10.0$ kpc), the
distribution is essentially flat (e.g. for iron the gradient is
d[Fe/H]/dR$_{\rm G} \approx -0.03$ dex/kpc).

As discussed in Paper~I and Paper~II, such a bimodal character in the
distribution may result from the combined effect of the radial gas flow
in the disc and the radial distribution of the star formation rate.
We note here that there are conflicting models of the galactic
structure, and that possibly the metallicity gradients can help to
decide which are the more likely ones. According to Sevenster
(\cite{sev99a}, \cite{sev99b}) and others (see references in Paper I) the bar
extends its influence to a co-rotation radius at about 4--6 kpc. In contrast,
according to Amaral \& L\'epine (\cite{amle97}) and others, the spiral arms
extend from the Inner Lindblad Resonance which is at about 2.5 kpc, to the
Outer Lindblad Resonance, at about 12 kpc, and the co-rotation of the
spiral pattern is close to the solar galactic orbit. In the vicinity of
a bar we expect to see elongated orbits of stars, and consequently, a
small metallicity gradient. On the other hand, according to
L\'epine, Mishurov \& Dedikov (\cite{lmd01})
an interaction between the gas and spiral waves in the disc forces the
gas to flow in opposite directions inside and outside the Galactic
co-rotation annulus. This mechanism produces a cleaning effect
in the middle part of the disc and consequently a flattening of the
metallicity distribution. At the same time, a decreased star formation
rate  in the vicinity of the galactic co-rotation, where the
relative velocity of the
spiral arms and of the gas passing through these arms is small, should
also result in some decrease in the abundances.

As the next step of this study we have begun to investigate the radial
abundance distribution in the outer disc. The region of primary
interest is at a galactocentric radius R$_{\rm G} \approx 10$ kpc, where
according to Twarog et al. (\cite{twaet97}) there exists a discontinuity
in the metallicity distribution.
Such a discontinuity can be suspected from earlier works of Janes
(\cite{jan79}), Panagia \& Tosi (\cite{pantos81}), and Friel (\cite{fri95}).
However, Twarog et al. (\cite{twaet97}) were the first to clearly stress this
result. Twarog et al. used photometic metallicities (interpreted to imply
[Fe/H]) for a large sample of open clusters, and they found that galactic
disc breaks into two distinct zones. Between R$_{\rm G} \approx 6.5-10.0$ kpc
they found a mean iron abundance $<$[Fe/H]$>$ of $\approx 0$
(i.e., the slope is very small, if present). Beyond R$_{\rm G} \approx 10.0$ kpc
the mean $<$[Fe/H]$>$ is $\approx-0.3$. This implies a sharp discontinuity
at R$_{\rm G} \approx 10$ kpc.

Recently, Caputo et al. (\cite{capet01}) reported a similar result. Those authors
calibrated BVI data for a large sample of galactic Cepheids (galactocentric distances
from 6 to 19 kpc) as a function of metallicity using non-linear pulsation
models. Their results (although not very reliable on a per star basis) suggest that
the derived metallicity distribution in the galactic disc can be represented either by
a single gradient of $-0.05$ dex~kpc$^{-1}$, or by a two-zone distribution
with a slope of $-0.01 \pm 0.05$ dex~kpc$^{-1}$ within 10 kpc and
$-0.02 \pm 0.02$ dex~kpc$^{-1}$ in the outer region of the galactic disc.
In other words, within each region the metallicity gradient is weak
to non-existent, while between these regions a significant change of the
metallicity/gradient does occur.

These results require an independent check, as well as further
specification of the quantitative characteristics of the radial abundance
distribution in the outer part of the galactic disc. With this aim we have organized a
separate study of distant Cepheids covering the span of distances
from 10 to 12 kpc. A short description of the observational material is given
in the next section.

\section{Observations}

Observations were carried out between 29 November and 2 December of
2001 with the 1.52-m telescope at La Silla (ESO, Chile).
The echelle spectrograph FEROS equipped with a 2048x4096 CCD
was used. The resolving power was 48000. The maximum exposure duration was
60 min. All the spectra have a S/N ratio of about 100. Only for EE Mon
is the S/N lower ($\approx 50$ or less). Table 1 contains
some information about our program Cepheids.
The columns give the star name, the period P (days), the V magnitude,
the exposure times (s), JD, phase, and derived quantities, such as
the effective temperature, gravity and microturbulent velocity
(see next section).

\begin{table*}
\caption[]{Program stars: details of the observations and atmosphere
parameters}
\begin{tabular}{rlcrcccccc}
\hline
Star & P, day & V &  Exp. (s) & JD 2452240+&$\phi$ &T$_{\rm eff}$,\,K&
$\log$~g&V$_{\rm t}$,\, km~s$^{-1}$ \\
\hline
CS Ori    &  3.88939 & 11.381 &3600&2.72917&0.679& 6539& 2.5& 3.6\\
AE Tau    &  3.89645 & 11.679 &2400&2.65903&0.377& 5981& 2.3& 4.0\\
V504 Mon(s)&  3.907  & 11.814 &3600&3.74583&0.714& 6370& 2.0& 3.5\\
AA Mon    &  3.93816 & 12.707 &2400&3.70556&0.106& 6261& 2.1& 4.0\\
EK Mon    &  3.95794 & 11.048 &2400&3.63125&0.918& 6001& 2.0& 4.5\\
V495 Mon  &  4.096583& 12.427 &2400&4.83681&0.622& 5591& 1.6& 4.0\\
V508 Mon  &  4.133608& 10.518 &2400&2.76875&0.742& 5566& 1.6& 4.0\\
VW Pup    &  4.28537 & 11.365 &3600&4.66250&0.617& 5586& 1.9& 4.6\\
EE Mon    &  4.80896 & 12.941 &3600&5.66180&0.209& 6134& 1.8& 4.5\\
XX Mon    &  5.456473& 11.898 &3600&4.79792&0.704& 5533& 1.7& 4.2\\
WW Pup    &  5.516724& 10.554 &3000&4.70833&0.722& 5550& 1.6& 4.2\\
TZ Mon    &  7.428014& 10.761 &2400&3.66944&0.817& 5008& 2.2& 1.5\\
AC Mon    &  8.01425 & 10.067 &2000&5.74167&0.081& 6075& 1.9& 4.8\\
TX Mon    &  8.70173 & 10.960 &2000&2.80417&0.122& 5871& 1.8& 4.9\\
HW Pup    & 13.454   & 12.050 &3600&4.75000&0.389& 5787& 1.3& 3.8\\
AD Pup    & 13.5940  &  9.863 &2400&3.78611&0.832& 5124& 0.9& 5.1\\
BN Pup    & 13.6731  &  9.882 &2400&3.81944&0.396& 5105& 1.0& 3.5\\
SV Mon    & 15.23278 &  8.219 &1800&2.69236&0.559& 4924& 1.1& 4.5\\
VZ Pup    & 23.1710  &  9.621 &1800&2.81944&0.980& 6070& 2.2& 4.7\\
\hline
\end{tabular}
\begin{list}{}{}
\item[] For V504 Mon (classified as s-Cepheid) the fundamentalized period
is given (P$_{0}$=P$_{1}$/0.71), P(observ)=P$_{1}$=2.77405.
\end{list}
\end{table*}

\section{Atmosphere parameters, abundances and distances}

In Paper~I the pertinent details concerning the determination
of atmospheric parameters and elemental abundances can be found along with
a description of our method of estimating the galactocentric distances.
The same methods were used in Paper~II and here.

To estimate the heliocentric distances of program Cepheids we used the
"absolute magnitude - pulsational period" relation of Gieren, Fouqu\'e
\& G\'omez (\cite{gfg98}). E(B-V), $<$B-V$>$, mean visual magnitudes and
pulsational periods are from Fernie et al. (\cite{feret95}), see Table 4.
We use for the interstellar absorption A$_{\rm v}$ an expression from Laney
\& Stobie (\cite{ls93}).

For s-Cepheids (DCEPS type) the observed periods are those of the first
overtone (see, e.g. Christensen-Dalsgaard \& Petersen \cite{chdp95}). Therefore,
following Paper~I and Paper~II for these stars the corresponding
periods of the unexcited fundamental mode were found using the ratio
P$_{1}$/P$_{0} \approx 0.72$, and these periods were then used to estimate
the absolute magnitudes.

In Tables 1--4 we list the adopted atmospheric parameters of our program stars,
derived elemental abundances and distance estimates, together with other
(hopefully) useful data. Note that the abundances derived from the lines
of different ions are given in the Appendix (Table A1) along with statistical
information about the abundances. The abundances of some elements (e.g. N, Sr,
Ru, Pr) listed in the Appendix were not used in the present gradient consideration,
since the corresponding abundances of these elements were derived only
for a minority of the program stars from Paper~I and Paper~II.

\begin{table*}
\caption[]{Averaged relative-to-solar elemental abundance for program Cepheids:
C--Mn}
\begin{tabular}{lrrrrrrrrrrrrrr}
\hline
  Star       &   C  &   N  &   O  & Na   & Mg   & Al   &Si    &  S   & Ca   & Sc   & Ti   & V    & Cr   & Mn   \\
\hline
CS Ori       &--0.58&--0.12&--0.61&  0.05&--0.40&--0.31&--0.16&--0.32&--0.20&--0.11&--0.23&  --  &--0.24&--0.43\\
AE Tau       &--0.48&  0.07&--0.17&--0.10&--0.36&--0.30&--0.18&--0.33&--0.25&--0.08&--0.06&--0.21&--0.12&--0.41\\
V504 Mon     &--0.53&  0.14&--0.35&--0.08&--0.45&--0.13&--0.17&--0.23&--0.21&--0.33&--0.28&--0.11&--0.30&--0.38\\
AA Mon       &--0.49&  0.35&--0.19&  0.04&--0.22&--0.14&--0.12&--0.12&--0.26&--0.21&--0.23&  0.02&--0.26&--0.29\\
EK Mon       &--0.42&  0.26&--0.28&  0.04&--0.18&--0.04&--0.05&--0.15&--0.10&--0.05&--0.18&--0.10&--0.15&--0.19\\
V495 Mon     &--0.77&--0.12&--0.09&  0.00&--0.28&--0.25&--0.18&--0.07&--0.29&--0.34&--0.29&--0.28&--0.32&--0.39\\
V508 Mon     &--0.77&  0.07&--0.54&--0.01&--0.28&--0.15&--0.18&--0.19&--0.22&--0.21&--0.13&--0.22&--0.22&--0.41\\
VW Pup       &--0.48&--0.12&  0.19&  0.02&--0.31&--0.21&--0.16&--0.29&--0.23&--0.19&--0.28&--0.31&--0.07&--0.37\\
EE Mon       &--0.48&  0.14&  --  &--0.20&--0.60&  --  &--0.35&--0.56&--0.50&--0.42&--0.26&--0.41&--0.35&--0.66\\
XX Mon       &--0.52&  0.15&  0.06&  0.02&  --  &  0.13&--0.18&--0.21&--0.18&--0.10&--0.23&--0.06&--0.13&--0.29\\
WW Pup       &--0.64&--0.16&  --  &  0.01&--0.33&--0.12&--0.14&--0.20&--0.22&--0.31&--0.25&--0.19&--0.20&--0.31\\
TZ Mon       &--0.53&  0.26&  0.28&  0.15&--0.05&  0.05&  0.00&  0.20&  0.06&--0.27&--0.08&--0.08&--0.05&  0.41\\
AC Mon       &--0.52&  0.23&--0.25&  0.05&  --  &--0.36&--0.08&--0.21&--0.13&--0.21&--0.19&--0.23&--0.24&--0.25\\
TX Mon       &--0.41&  0.29&--0.16&--0.08&--0.26&--0.07&--0.09&--0.17&--0.24&--0.16&--0.17&--0.28&--0.16&--0.25\\
HW Pup       &--0.56&  0.04&--0.29&--0.20&--0.52&--0.16&--0.19&--0.29&--0.27&--0.30&--0.25&--0.33&--0.29&--0.28\\
AD Pup       &--0.77&  0.49&  --  &  0.06&  --  &  0.00&--0.19&--0.13&--0.17&--0.04&--0.22&--0.32&--0.38&--0.33\\
BN Pup       &--0.45&  0.55&--0.06&  0.09&  0.21&  0.04&--0.01&  0.19&  0.04&--0.02&--0.05&--0.15&  0.07&--0.02\\
SV Mon       &--0.81&  0.22&  0.21&  0.10&--0.48&--0.04&--0.02&  0.24&--0.13&--0.07&--0.10&--0.19&--0.05&  0.03\\
VZ Pup       &--0.83&  0.02&--0.10&  0.00&--0.15&--0.15&--0.09&--0.22&--0.12&--0.16&--0.26&--0.01&--0.18&--0.28\\
\hline
\end{tabular}
\end{table*}

\begin{table*}
\caption[]{Same as Table 2 but for Fe--Gd}
\begin{tabular}{lrrrrrrrrrrrr}
\hline
  Star       & Fe   & Co   & Ni   & Cu   & Zn   & Y    & Zr   & La   & Ce   & Nd   & Eu   & Gd   \\
\hline
CS Ori       &--0.26&  --  &--0.25&  0.19&  --  &--0.09&  --  &  0.21&  0.07&--0.17&--0.03&  --  \\
AE Tau       &--0.19&--0.45&--0.22&--0.14&  --  &  0.08&--0.18&  0.32&--0.27&--0.08&  0.11&  --  \\
V504 Mon     &--0.31&  --  &--0.36&  0.20&  --  &--0.17&--0.24&  --  &--0.26&--0.30&--0.15&  --  \\
AA Mon       &--0.21&  --  &--0.30&  0.29&  --  &  0.07&  --  &  --  &  0.03&--0.11&--0.32&  --  \\
EK Mon       &--0.10&--0.06&--0.20&--0.96&  --  &--0.07&  0.05&  --  &--0.22&  0.06&  0.31&  --  \\
V495 Mon     &--0.26&--0.25&--0.36&--0.31&  --  &--0.10&--0.44&  0.05&  0.04&--0.14&--0.17&  --  \\
V508 Mon     &--0.25&--0.35&--0.35&--0.30&  0.11&--0.17&--0.16&  0.06&--0.17&--0.24&--0.27&--0.14\\
VW Pup       &--0.19&--0.26&--0.17&--0.64&  --  &--0.09&  --  &  0.19&  0.08&--0.06&--0.08&  --  \\
EE Mon       &--0.51&  --  &--0.58&  --  &  --  &--0.40&  --  &  --  &--0.27&--0.23&  --  &  --  \\
XX Mon       &--0.18&--0.27&--0.27&--0.22&  --  &  0.00&--0.25&  0.06&--0.05&  0.01&  0.04&  --  \\
WW Pup       &--0.18&  --  &--0.21&--0.34&--0.01&  0.04&  --  &  0.15&  0.00&--0.04&--0.09&  --  \\
TZ Mon       &--0.03&--0.16&--0.09&  0.17&  0.26&--0.21&--0.22&  0.04&--0.26&--0.25&--0.01&  --  \\
AC Mon       &--0.22&  --  &--0.35&--0.80&  --  &--0.12&  --  &  0.13&--0.20&--0.12&  0.18&  --  \\
TX Mon       &--0.14&--0.24&--0.20&--0.04&  0.15&  0.06&--0.27&  0.23&  0.10&--0.10&  0.07&--0.23\\
HW Pup       &--0.29&--0.18&--0.35&--0.32&  --  &--0.13&  --  &  --  &--0.13&--0.15&--0.11&  --  \\
AD Pup       &--0.24&--0.48&--0.37&--0.47&  --  &--0.11&--0.20&  0.27&--0.14&--0.12&--0.06&--0.03\\
BN Pup       &  0.01&--0.23&--0.03&--0.24&  --  &  0.23&--0.17&  0.30&  0.00&  0.03&  0.07&  --  \\
SV Mon       &  0.00&--0.24&--0.08&--0.42&  --  &  0.17&  0.00&  0.30&  0.08&  0.16&  0.08&  0.12\\
VZ Pup       &--0.16&  --  &--0.28&  0.10&  --  &--0.06&  --  &  --  &--0.11&  0.02&  0.02&  0.32\\
\hline
\end{tabular}
\end{table*}

\begin{table*}
\caption[]{Physical and positional characteristics for program Cepheids}
\begin{tabular}{rrcrccrrrr}
\hline
Star      & P, d  &$<$(B-V)$>$&E(B-V)&M$_{\rm v}$&d, pc& l & b &R$_{\rm G}$, kpc & [Fe/H] \\
\hline
CS Ori    &  3.88939 &0.924 &0.402 &--2.93&3997.6&197.97&--4.51&11.76&--0.26\\
AE Tau    &  3.89645 &1.129 &0.604 &--2.93&3389.8&181.03&--3.52&11.28&--0.19\\
V504 Mon  &  3.907   &1.036 &0.565 &--2.93&3845.9&215.69&--3.29&11.24&--0.31\\
AA Mon    &  3.93816 &1.409 &0.832 &--2.94&3860.2&217.03&--0.43&11.22&--0.21\\
EK Mon    &  3.95794 &1.195 &0.584 &--2.95&2618.1&215.27&--0.82&10.15&--0.10\\
V495 Mon  &  4.096583&1.241 &0.640 &--2.99&4630.0&213.81&--4.49&12.01&--0.26\\
V508 Mon  &  4.133608&0.898 &0.323 &--3.00&3120.1&208.91&  0.86&10.74&--0.25\\
VW Pup    &  4.28537 &1.065 &0.514 &--3.04&3533.7&235.37&--0.62&10.33&--0.19\\
EE Mon    &  4.80896 &0.966 &0.488 &--3.18&8130.6&219.97&--3.77&15.05&--0.51\\
XX Mon    &  5.456473&1.139 &0.596 &--3.33&4566.6&215.52&--1.11&11.92&--0.18\\
WW Pup    &  5.516724&0.874 &0.398 &--3.35&3342.7&237.38&  0.97&10.10&--0.18\\
TZ Mon    &  7.428014&1.116 &0.441 &--3.71&4020.6&214.01&  1.29&11.45&--0.03\\
AC Mon    &  8.01425 &1.165 &0.508 &--3.80&2754.8&221.77&--1.85&10.12&--0.22\\
TX Mon    &  8.70173 &1.096 &0.511 &--3.90&4350.8&214.14&--0.78&11.76&--0.14\\
HW Pup    & 13.454   &1.237 &0.723 &--4.42&6684.1&244.77&  0.78&12.33&--0.29\\
AD Pup    & 13.5940  &1.049 &0.330 &--4.43&4388.0&241.93&--0.03&10.69&--0.24\\
BN Pup    & 13.6731  &1.186 &0.438 &--4.44&3762.1&247.90&  1.07& 9.95&  0.01\\
SV Mon    & 15.23278 &1.048 &0.249 &--4.57&2472.3&203.74&--3.67&10.21&  0.00\\
VZ Pup    & 23.1710  &1.162 &0.471 &--5.07&4263.0&243.42&--3.31&10.52&--0.16\\
\hline
\end{tabular}
\end{table*}

\section{Radial abundance distributions: from inner to outer disc}

Using the results presented in Tables 2--4 one can construct the radial
abundance distribution in the outer disc. To make the picture on galactic
abundance gradients as complete as possible, we plotted data from the present
paper together with those from Paper~I and Paper~II  (for the sake of
clarity the error bars are shown only for the new sample). Figs. 1--5
display the derived dependencies between the abundances of 25 chemical
elements and galactocentric distances. As the iron abundances are the most
reliable we will concentrate our discussion on the iron gradient
d[Fe/H]/dR$_{\rm G}$.

\begin{figure}
\resizebox{\hsize}{!}{\includegraphics{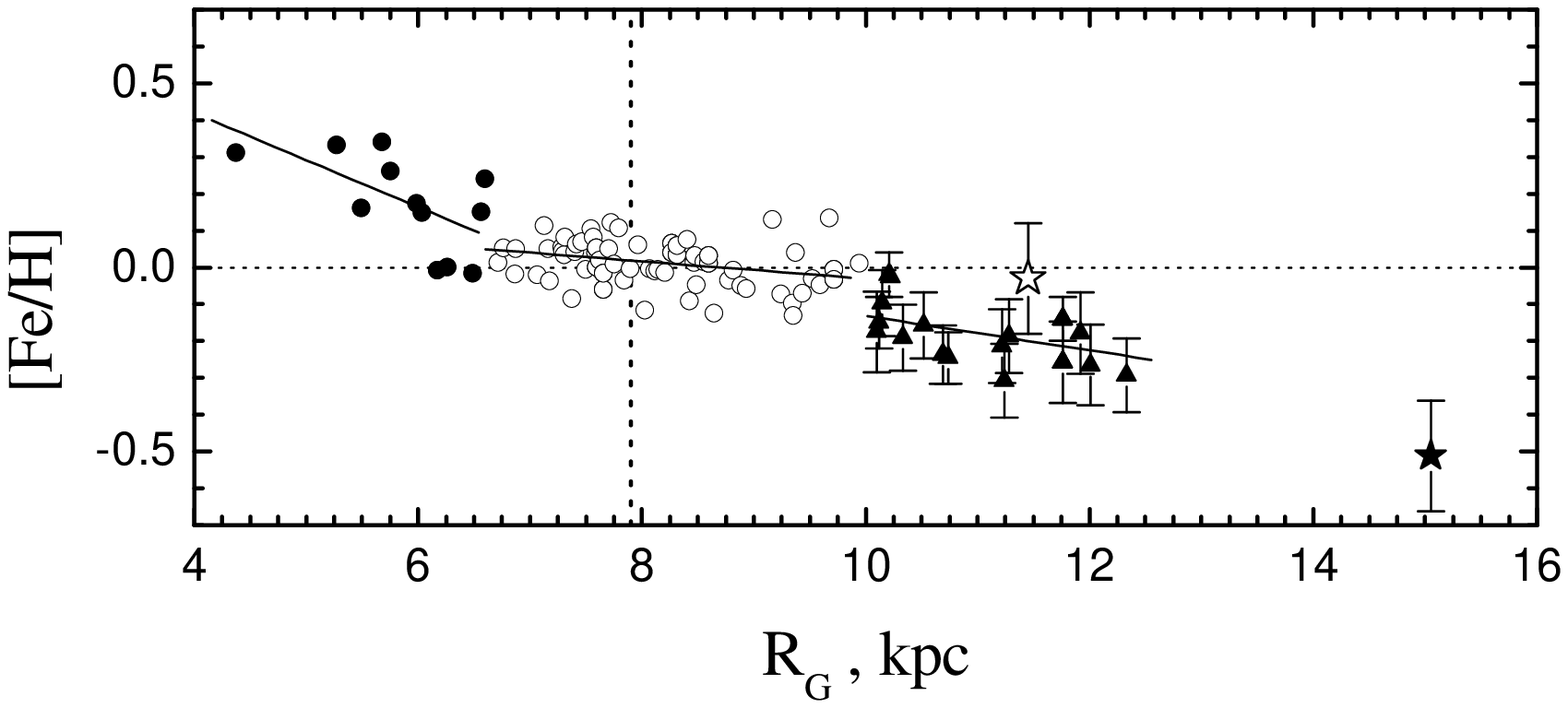}}
\caption[]{The radial distribution of the iron abundance. $Open~circles$ -
the data from Paper~I, $black~circles$ - the data from Paper~II,
$black~triangles$ - present results. 1-$\sigma$ interval is indicated for
the stars from the present study (the error of the mean is of the order of
0.01).
The position of EE~Mon is indicated by $filled~asterisk$, and TZ Mon --
by $open~asterisk$. The Sun is marked by the intersection of the
dashed lines.}
\end{figure}

\newpage

\begin{figure*}
\resizebox{\hsize}{!}{\includegraphics{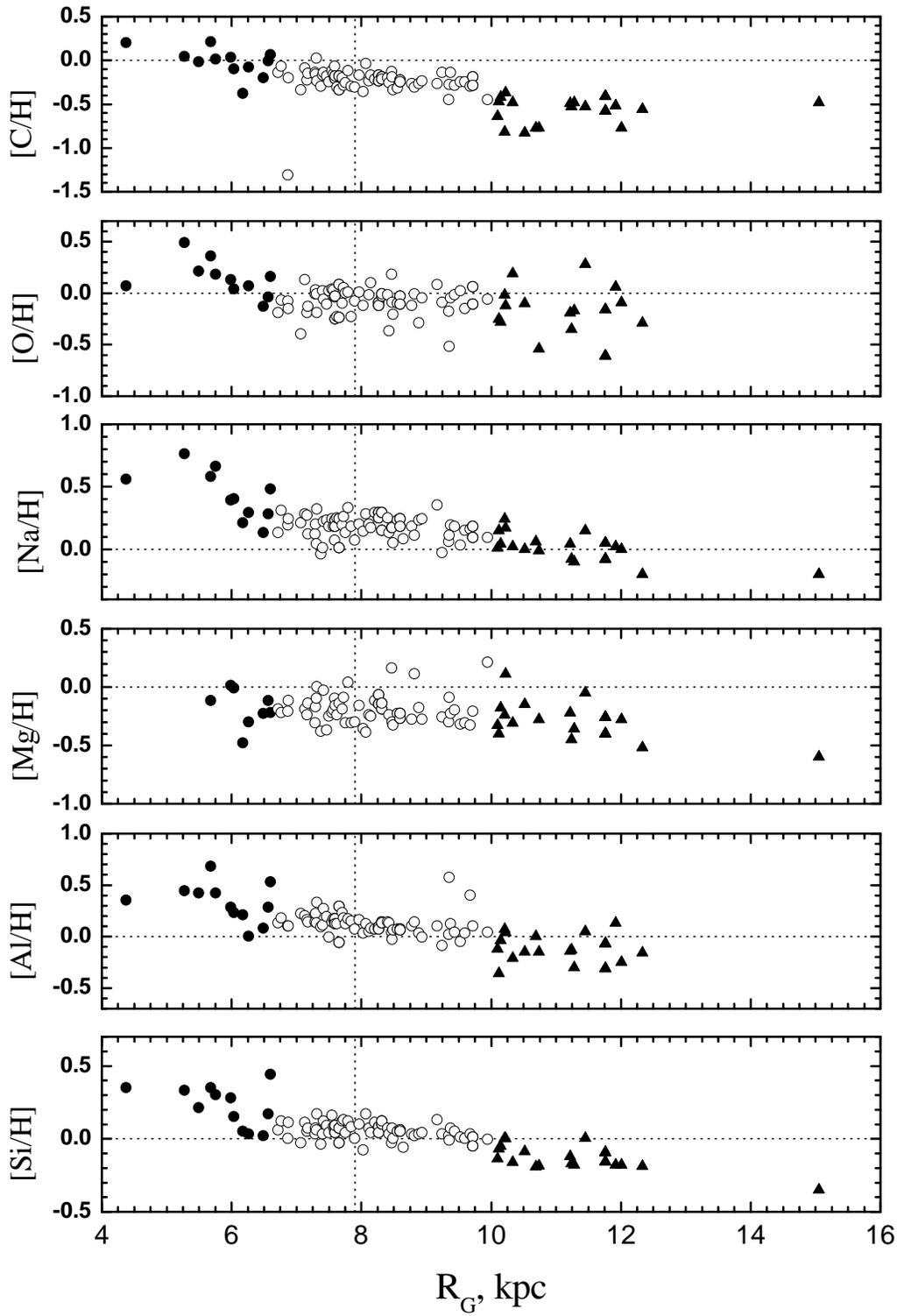}}
\caption[]{Same as Fig. 1, but for elements C--Si}
\end{figure*}

\newpage

\begin{figure*}
\resizebox{\hsize}{!}{\includegraphics{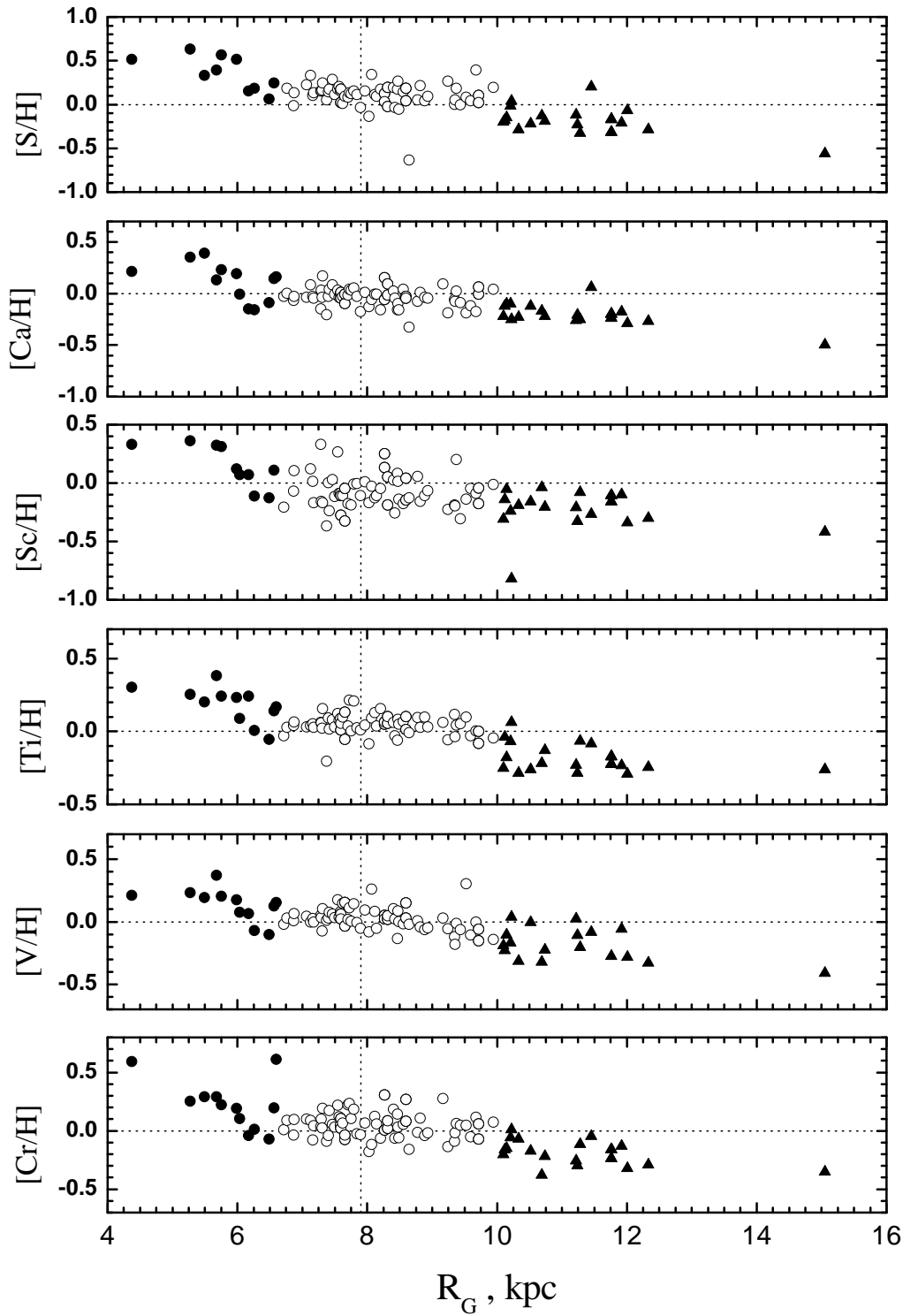}}
\caption[]{Same as Fig. 1, but for elements S--Cr}
\end{figure*}

\newpage

\begin{figure*}
\resizebox{\hsize}{!}{\includegraphics{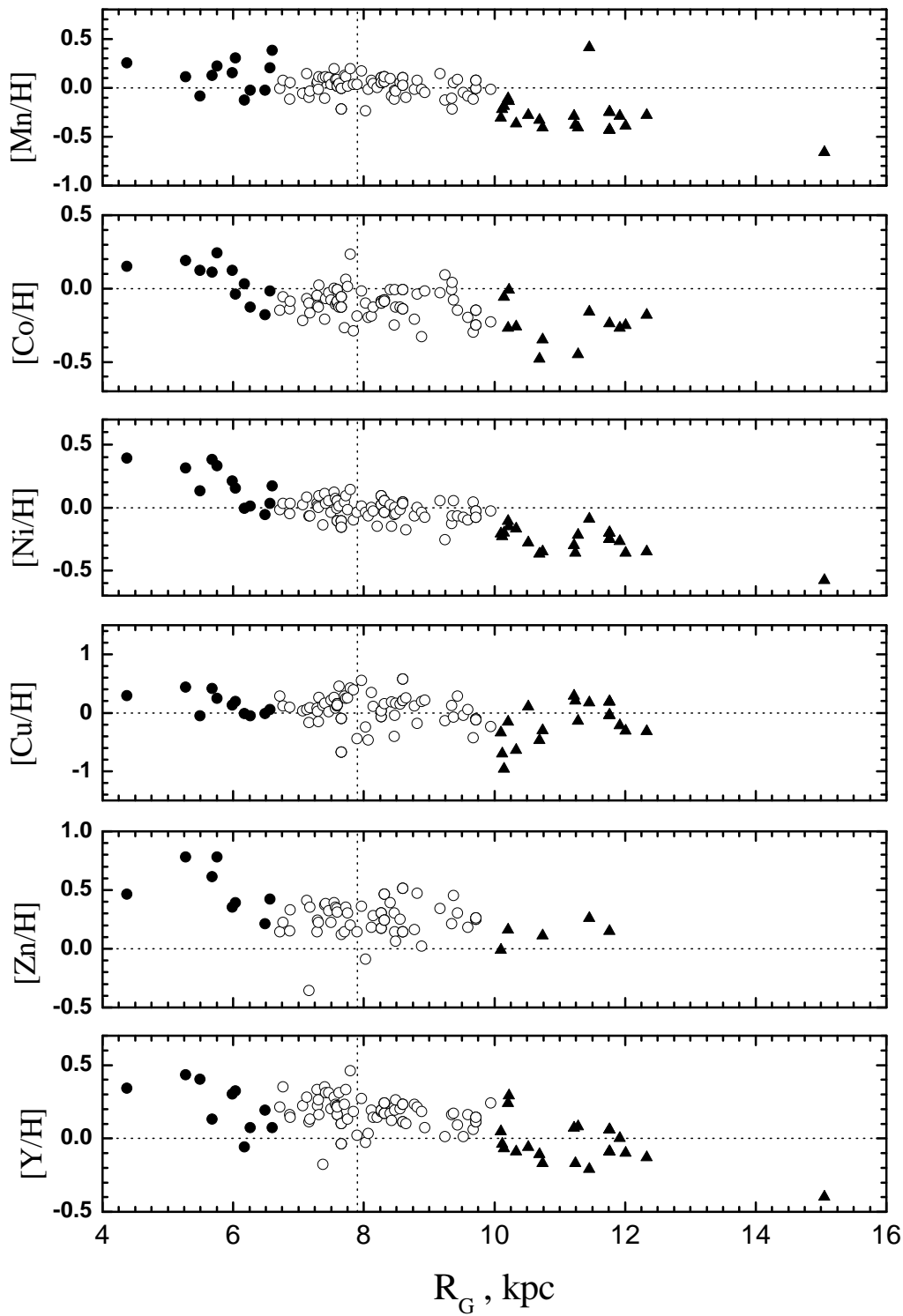}}
\caption[]{Same as Fig. 1, but for elements Mn--Y}
\end{figure*}

\newpage

\begin{figure*}
\resizebox{\hsize}{!}{\includegraphics{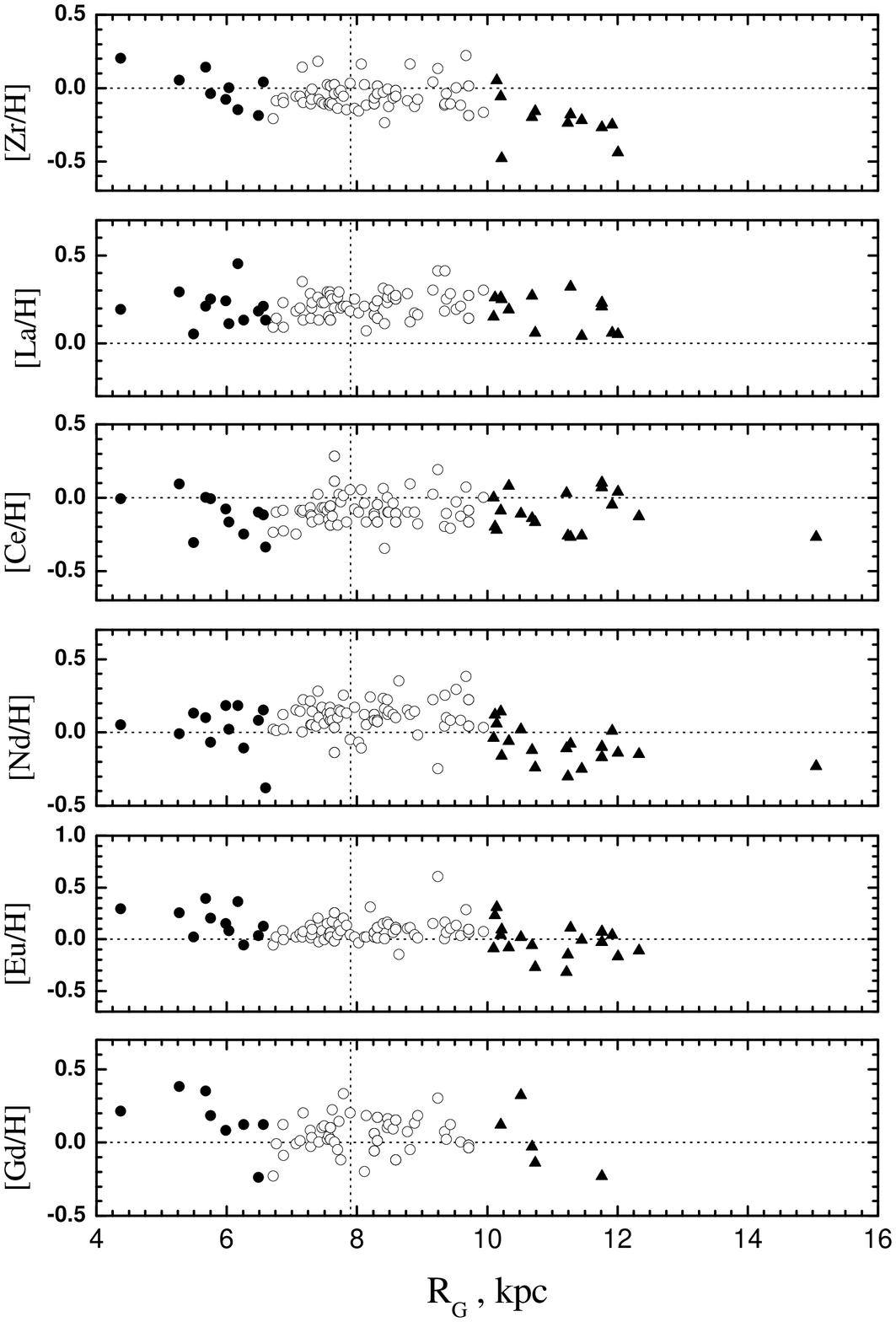}}
\caption[]{Same as Fig. 1, but for elements Zr--Gd}
\end{figure*}

\section{Discussion}

\subsection{Iron abundance gradient}

The specific aim of this paper is to extend the current work to
greater distances towards the galactic anticenter, and to search for
possible irregularities in the abundance distributions.
Distances can be separated into three zones: The inner part of the galactic
disk (gradient d[Fe/H]/dR$_{\rm G}
\approx -0.13\pm 0.03$ dex~kpc$^{-1} $), the mid part of the disk (gradient
$\approx -0.02\pm 0.01$ dex~kpc$^{-1} $), and a piece of outer disc.
For the latter we derive a gradient $-0.06\pm 0.01$ dex~kpc$^{-1}$
and a mean [Fe/H] $\approx -0.19\pm 0.08$ dex. The gradient
for each zone was derived from a least-squares fit using the weighted
data (for the weights assigned to the stars of the inner and mid
parts of the disc see Paper~I and Paper~II). Each star of the present
sample was assigned unit weight (W = 1). Two stars (TV Cam and YZ Aur) from
the outer zone were excluded from the present determination. For those
Cepheids we have analyzed only photographic spectra (see Paper~I),
therefore we consider the abundance results (only iron) for TV Cam and
YZ Aur as being much less reliable than for other stars where we have the
CCD spectra. We also excluded from the statistics the very distant Cepheid
EE Mon, and the anomalous Cepheid TZ Mon (about the latter see below).
The former will be included in the statistics consideration after we
sample the region of galactocentric distances from 12 kpc to 15 kpc
(next papers of this series).

The transition zone at 10 kpc can be easily identified in Fig. 1. After this
point the metallicity drops by approximately 0.2 dex. All the stars in the bin
beyond 10 kpc are iron-deficient. The same result is seen in Fig.3ab of
Twarog et al. (\cite{twaet97}) which shows their open cluster metallicity
values as a function of galactocentric radius. It should be noted that the
sample of open clusters used by Twarog et al. consists of the clusters with
ages spanning from 1 to 5 Gyr.
Thus, the youngest clusters used for the gradient study are
approximately 10 times older than Cepheids. By comparing the iron abundance
gradient from Cepheids with that from open clusters one would, in principle,
estimate how the abundance gradient evolved with time. Nevertheless, in
practice it is difficult to realize, because any conclusion will suffer from
the rather high uncertainty of the open cluster data. The only we can state
confidently is that the discontinuity of the metallicity distribution has
really survived over several Gyrs, until now.

Using a linear fit to the [Fe/H] data over the entire baseline (4--12 kpc)
produces a gradient of $-0.06$ dex~kpc$^{-1}$ which is surprisingly close
to the canonical literature value of $-0.06/-0.07$ dex~kpc$^{-1}$
(see Papers I and II and references therein).

\subsection{Radial distribution of other species}

In this paper we will not discuss in detail the radial distributions of the
elements other than iron. This will be done in a future paper using the whole
observed baseline 4--15 kpc after the region of galactocentric distances
between 12 and 15 kpc is well sampled.

As is well known, the carbon abundance is altered during
the Cepheid evolution, specifically as a result of the first dredge-up.
Therefore we do not attach a great significance to the radial distribution
obtained for this element. Nevertheless, interestingly enough,
the [C/H]--R$_{\rm G}$ dependence is rather clear.
Two remarkable features inherent to carbon abundance distribution should
be stressed. First, the extremely low carbon content in FN Aql (see Paper~I
and Fig. 2, R$_{\rm G} \approx 6.9$ kpc), and second, the well expressed
"dip" in carbon abundance at galactocentric distances near 10 kpc.
The presence of a coherent behavior in carbon can be interpreted as the
original carbon abundances show a similar gradient, and as the mixing
event at the first dredge-up modifies all of the abundances by similar
amounts; i.e., that the mixing process has similar efficiency (depth)
for all stars in this mass range.

We believe that rather large scatter in the radial distribution of oxygen
abundance at distances 10--12 kpc is probably caused by some unreliability
in oxygen abundance derived for distant Cepheids. This scatter probably does
not reflect the real inhomogeneity of oxygen abundance in the interstellar
media of the outer disc. The same can be said about elements such as Sc, Cu, Zn
(only a few spectral lines for each of these elements including O were analyzed).
Those elements whose abundances were based on the analysis of a
significant number of lines (e.g. Si, Ti, V, Ni), show rather tight
distributions.

The step distribution and apparent discontinuity (similar to iron) is
also seen for Si, Ti, Mn, Co, Ni, Zr and Nd.

\subsection{A possible explanation for the observed features}

If we consider the ensemble of our results, paying special attention to the
elements for which the data present smaller scatter because they present a
large number of lines in the spectra, like Si, Ti, V, Ni, in addition to Fe
and C, it is very clear that the metallicity distribution is flat between
about 6 and 10 kpc, and that there is a discontinuity of about $-0.2$ dex
for R$_{\rm G} > 10$ kpc, and perhaps a minimum at about 10.5--11.0 kpc which
can be traced from some distributions (let us recall that we adopt
R$_{\odot} = 7.9$ kpc). This is a new result, although the discontinuity
at about 10 kpc was already observed by Twarog et al. (\cite{twaet97}), in
a sample of open clusters. An attractive explanation for the minimum at
10.5 kpc (if it truly exists) is the minimum of star formation rate that
is expected at co-rotation, as mentioned in the Introduction.

Recently Mishurov et al. (\cite{mishet02}) produced a detailed model of
chemical evolution of the galactic disc, taking into account the effect of
co-rotation, to explain our data presented in Papers I and II. The new data
presented in this paper, suggest that the model of Mishurov et al.
(\cite{mishet02}) is basically correct, but that the co-rotation radius should
be slightly shifted to about 10.5 kpc. The discontinuity in the metallicity
distribution at 10 kpc is possibly explained by the gap in the gas density
distribution that is associated with co-rotation (see L\'epine, Mishurov \&
Dedikov \cite{lmd01}). If we divide the Galactic disc in a large number of
concentric rings, the gas from neighbouring rings tends to mix due to
supernova explosions, stellar winds, cloud collisions, etc., that do not
respect the frontiers between concentric rings. This mixing is equivalent
to a diffusion term, and tends to smooth out metallicity gradients in the
gas (and therefore, in recently formed stars). However, the gas density gap
associated with co-rotation, which is observed in the 21 cm hydrogen line as
discussed by L\'epine, Mishurov \& Dedikov (\cite{lmd01}) is possibly a
barrier that avoids contact between the gas at R$_{\rm G} >$ R$_{\rm c}$ and
R$_{\rm G} <$ R$_{\rm c}$ and allows the existence of two distinct zones.


\subsection{Some notes about TZ Mon}

This star differs from other Cepheids by its very low V$_{\rm t}$ value
(1.5~km~s$^{-1}$, see Table 1) which is not appropriate for supergiants.
Nevertheless, only by using this small value can one avoid any dependence
between iron abundance from individual iron lines and their equivalent
widths. With the adopted microturbulent velocity this distant Cepheid shows
solar-like abundances for the great majority of elements. For Mn we
detected a strong overabundance. Another interesting feature, which is
seen in the TZ Mon spectrum, is the 6707 \AA~Li~I line. Normally this
line is not present in supergiant spectra. The equivalent width of
W=10 m\AA~ results in an absolute lithium abundance for TZ Mon of about
0.84.  This is consistent with Li abundances previously determined for some
non-variable G/K supergiants by Luck (\cite{luck77}).

\begin{acknowledgements}
This work was supported in part by a grant to SMA from FAPESP
S\~ao Paulo (visiting professor fellowship No. 2000/06587-3),
and ESO Garching (visitor programme). SMA is very grateful to
Drs. B Barbuy, M. Spite, F. Spite and S. D'Odorico for their personal
assistance. He would like also to thank the staff of ESO La Silla
for the provision of excellent conditions during his observations.

The authors are indebted to the referee, Dr. B.W. Carney, for important comments
which allowed us to improve the fist version of this paper.
\end{acknowledgements}

{}

\newpage

\begin{appendix}

{\bf Appendix}
{\bf Table A1} Elemental abundances in program Cepheids\\
\tiny
\begin{tabular}{lrrrrrrrrrrrrrrrr}
\hline
\noalign{\smallskip}
\multicolumn{1}{c}{}&\multicolumn{4}{c}{CS Ori, P=3.88939}&
\multicolumn{4}{c}{AE Tau 3.89645}&\multicolumn{4}{c}{V504 Mon (s) 3.907}&
\multicolumn{4}{c}{AA  Mon 3.93816}\\
\noalign{\smallskip}
\hline
 Ion &[M/H]
&$\sigma$&N&(M/H)&[M/H]&$\sigma$&N&(M/H)&[M/H]&$\sigma$&N&(M/H)&[M/H]&$\sigma$&N&(M/H)\\
\hline
C I   &--0.58&  0.18&   5&7.97&--0.48 & 0.08 &  5 &8.07 &--0.53&  0.12&   9 &8.02 &--0.49 & 0.13&   6 &8.06\\
N I   &--0.12&  0.10&   8&7.85&  0.07 & 0.08 &  6 &8.04 &  0.14&  0.05&   7 &8.11 &  0.35 & 0.06&   5 &8.32\\
O I   &--0.61&   -- &   1&8.26&--0.17 & 0.23 &  2 &8.70 &--0.35&   -- &   1 &8.52 &--0.19 & 0.07&   2 &8.68\\
Na I  &  0.05&  0.01&   3&6.38&--0.10 & 0.01 &  2 &6.23 &--0.08&  0.10&   4 &6.25 &  0.04 & 0.13&   2 &6.37\\
Mg I  &--0.40&   -- &   1&7.18&--0.36 & 0.18 &  2 &7.22 &--0.45&   -- &   1 &7.13 &--0.22 & 0.12&   2 &7.36\\
Al I  &--0.31&  0.33&   2&6.16&--0.30 & 0.11 &  6 &6.17 &--0.13&   -- &   1 &6.34 &--0.14 &  -- &   1 &6.33\\
Si I  &--0.16&  0.11&  22&7.39&--0.18 & 0.10 & 23 &7.37 &--0.19&  0.13&  13 &7.36 &--0.15 & 0.18&   6 &7.40\\
Si II &--0.09&   -- &   1&7.46&--0.19 &  --  &  1 &7.36 &  0.10&   -- &   1 &7.65 &  0.06 &  -- &   1 &7.61\\
S I   &--0.32&  0.16&   5&6.89&--0.33 & 0.15 &  3 &6.88 &--0.23&  0.22&   4 &6.98 &--0.12 & 0.20&   5 &7.09\\
Ca I  &--0.20&  0.10&   9&6.16&--0.25 & 0.08 &  9 &6.11 &--0.21&  0.12&  11 &6.15 &--0.26 & 0.15&   8 &6.10\\
Sc I  &  0.38&   -- &   1&3.55&  0.20 &  --  &  1 &3.37 &   -- &   -- &  -- & --  &   --  &  -- &  -- & -- \\
Sc II &--0.27&  0.16&   3&2.90&--0.15 & 0.17 &  4 &3.02 &--0.33&  0.18&   5 &2.84 &--0.21 & 0.15&   6 &2.96\\
Ti  I &--0.20&  0.05&   4&4.82&--0.07 & 0.14 &  9 &4.95 &--0.22&  0.07&   3 &4.80 &--0.21 & 0.10&   3 &4.81\\
Ti II &--0.33&   -- &   1&4.69&--0.02 &  --  &  1 &5.00 &--0.35&  0.12&   3 &4.67 &--0.25 & 0.05&   3 &4.77\\
V  I  &   -- &   -- &  --& -- &--0.16 & 0.20 &  5 &3.84 &--0.11&   -- &   1 &3.89 &  0.18 & 0.01&   2 &4.18\\
V II  &   -- &   -- &  --& -- &--0.32 & 0.05 &  2 &3.68 &   -- &   -- &  -- & --  &--0.29 &  -- &   1 &3.71\\
Cr  I &--0.03&   -- &   1&5.64&--0.13 & 0.14 &  3 &5.54 &   -- &   -- &  -- & --  &--0.23 &  -- &   1 &5.44\\
Cr II &--0.31&  0.13&   3&5.36&--0.10 & 0.07 &  3 &5.57 &--0.30&  0.12&   9 &5.37 &--0.26 & 0.13&   7 &5.41\\
Mn I  &--0.43&  0.04&   3&4.96&--0.41 & 0.10 &  5 &4.98 &--0.38&  0.13&   2 &5.01 &--0.29 & 0.05&   2 &5.10\\
Fe I  &--0.26&  0.11&  80&7.24&--0.19 & 0.10 &113 &7.31 &--0.31&  0.10&  57 &7.19 &--0.21 & 0.10&  48 &7.29\\
Fe II &--0.25&  0.07&  18&7.25&--0.17 & 0.09 & 22 &7.33 &--0.30&  0.11&  29 &7.20 &--0.22 & 0.09&  21 &7.28\\
Co I  &   -- &   -- &  --& -- &--0.45 &  --  &  1 &4.47 &   -- &   -- &  -- & --  &    -- &   --&   --&  --\\
Ni I  &--0.25&  0.17&  24&6.00&--0.22 & 0.07 & 26 &6.03 &--0.36&  0.10&  11 &5.89 &--0.30 & 0.11&   8 &5.95\\
Cu I  &  0.19&   -- &   1&4.40&--0.14 &  --  &  1 &4.07 &  0.20&   -- &   1 &4.41 &  0.29 &  -- &   1 &4.50\\
Y II  &--0.09&  0.13&   3&2.15&  0.08 & 0.04 &  3 &2.32 &--0.17&  0.11&   4 &2.07 &  0.07 & 0.10&   5 &2.31\\
Zr II &   -- &   -- &  --& -- &--0.18 &  --  &  1 &2.41 &--0.24&   -- &   1 &2.36 &    -- &   --&   --&  --\\
La II &  0.21&   -- &   1&1.43&  0.32 &  --  &  1 &1.54 &   -- &   -- &  -- & --  &    -- &   --&   --&  --\\
Ce II &  0.07&   -- &   1&1.62&--0.27 & 0.19 &  2 &1.28 &--0.26&  0.20&   3 &1.28 &  0.03 & 0.17&   2 &1.58\\
Pr II &   -- &   -- &  --& -- &   --  &  --  & -- & --  &--0.39&   -- &   1 &0.32 &    -- &   --&   --&  --\\
Nd II &--0.17&   -- &   1&1.33&--0.08 & 0.15 &  3 &1.42 &--0.30&  0.14&   3 &1.20 &--0.11 & 0.06&   3 &1.39\\
Eu II &--0.03&   -- &   1&0.48&  0.11 & 0.01 &  2 &0.62 &--0.15&   -- &   1 &0.36 &--0.32 &  -- &   1 &0.19\\
\hline
\hline
\multicolumn{1}{c}{}&\multicolumn{4}{c}{EK Mon 3.95794}&
\multicolumn{4}{c}{V495 Mon  4.09658}&\multicolumn{4}{c}{V508 Mon  4.133608}&
\multicolumn{4}{c}{VW Pup 4.28537}\\
\noalign{\smallskip}
\hline
 Ion &[M/H]&$\sigma$&N&(M/H)&[M/H]&$\sigma$&N&(M/H)&[M/H]&$\sigma$&N&(M/H)&[M/H]&$\sigma$&N&(M/H)\\
\hline
C I    & --0.42&  0.10 &    6 &   8.13 &       --0.77&  0.08 &    3 &   7.78&      --0.77&  0.08&     5 &   7.78 &      --0.48&  0.14 &    4 &   8.07 \\
N I    &   0.26&  0.07 &    5 &   8.23 &       --0.12&   --  &    1 &   7.85&        0.07&  0.08&     6 &   8.04 &      --0.12&   --  &    1 &   7.85 \\
O I    & --0.28&   --  &    1 &   8.59 &       --0.09&  0.04 &    2 &   8.78&      --0.54&   -- &     1 &   8.33 &        0.19&   --  &    1 &   9.06 \\
Na I   &   0.04&  0.01 &    2 &   6.37 &       --0.00&  0.06 &    3 &   6.33&      --0.01&  0.12&     2 &   6.32 &        0.02&  0.00 &    2 &   6.35 \\
Mg I   & --0.18&   --  &    1 &   7.40 &       --0.28&   --  &    1 &   7.30&      --0.28&   -- &     1 &   7.30 &      --0.31&   --  &    1 &   7.27 \\
Al I   & --0.04&  0.00 &    2 &   6.43 &       --0.25&  0.23 &    4 &   6.22&      --0.15&  0.19&     4 &   6.32 &      --0.21&  0.02 &    2 &   6.26 \\
Si I   & --0.05&  0.07 &   10 &   7.50 &       --0.18&  0.07 &   14 &   7.37&      --0.20&  0.07&    17 &   7.35 &      --0.17&  0.07 &   10 &   7.38 \\
Si I   &    -- &   --  &  --  &    --  &          -- &   --  &  --  &    -- &      --0.05&  0.13&     2 &   7.50 &      --0.12&  0.19 &    2 &   7.43 \\
S I    & --0.15&  0.03 &    3 &   7.06 &       --0.07&   --  &    1 &   7.14&      --0.19&  0.29&     4 &   7.02 &      --0.29&  0.20 &    2 &   6.92 \\
Ca I   & --0.10&  0.09 &    5 &   6.26 &       --0.29&  0.16 &    6 &   6.07&      --0.22&  0.07&     4 &   6.14 &      --0.23&  0.06 &    5 &   6.13 \\
Sc I   &   0.49&   --  &    1 &   3.66 &          -- &   --  &  --  &    -- &        0.08&   -- &     1 &   3.25 &         -- &   --  &  --  &    --  \\
Sc II  & --0.19&  0.19 &    4 &   2.98 &       --0.34&  0.11 &    5 &   2.83&      --0.35&  0.12&     2 &   2.82 &      --0.19&  0.12 &    3 &   2.98 \\
Ti  I  & --0.20&   --  &    1 &   4.82 &       --0.29&  0.10 &    5 &   4.73&      --0.10&  0.12&     7 &   4.92 &      --0.26&   --  &    1 &   4.76 \\
Ti II  & --0.17&  0.08 &    3 &   4.85 &       --0.30&  0.11 &    2 &   4.72&      --0.34&   -- &     1 &   4.68 &      --0.31&   --  &    1 &   4.71 \\
V  I   & --0.07&   --  &    1 &   3.93 &       --0.22&  0.20 &    3 &   3.78&      --0.20&  0.13&     7 &   3.80 &      --0.35&  0.08 &    2 &   3.65 \\
V II   & --0.14&   --  &    1 &   3.86 &       --0.47&   --  &    1 &   3.53&      --0.40&   -- &     1 &   3.60 &      --0.28&  0.09 &    2 &   3.72 \\
Cr  I  &    -- &   --  &  --  &    --  &       --0.21&   --  &    1 &   5.46&      --0.22&   -- &     1 &   5.45 &      --0.04&  0.29 &    2 &   5.63 \\
Cr II  & --0.15&  0.07 &    7 &   5.52 &       --0.34&  0.14 &    7 &   5.33&      --0.22&  0.08&     5 &   5.45 &      --0.09&  0.20 &    3 &   5.58 \\
Mn I   & --0.19&  0.08 &    3 &   5.20 &       --0.39&  0.10 &    6 &   5.00&      --0.41&  0.06&     4 &   4.98 &      --0.37&  0.08 &    4 &   5.02 \\
Fe I   & --0.09&  0.09 &   71 &   7.41 &       --0.26&  0.11 &   68 &   7.24&      --0.25&  0.07&    94 &   7.25 &      --0.19&  0.09 &   73 &   7.31 \\
Fe II  & --0.12&  0.08 &   21 &   7.38 &       --0.28&  0.12 &   21 &   7.22&      --0.23&  0.09&    22 &   7.27 &      --0.20&  0.08 &   12 &   7.30 \\
Co I   & --0.06&   --  &    1 &   4.86 &       --0.25&  0.07 &    3 &   4.67&      --0.35&  0.00&     2 &   4.57 &      --0.26&  0.06 &    4 &   4.66 \\
Ni I   & --0.20&  0.09 &   18 &   6.05 &       --0.36&  0.08 &   18 &   5.89&      --0.35&  0.07&    22 &   5.90 &      --0.17&  0.12 &   17 &   6.08 \\
Cu I   & --0.96&   --  &    1 &   3.25 &       --0.31&  0.04 &    2 &   3.90&      --0.30&   -- &     1 &   3.91 &      --0.64&  0.39 &    2 &   3.57 \\
Zn I   &    -- &   --  &  --  &    --  &          -- &   --  &  --  &    -- &        0.11&   -- &     1 &   4.71 &         -- &   --  &  --  &    --  \\
Y II   & --0.07&  0.08 &    6 &   2.17 &       --0.10&  0.09 &    6 &   2.14&      --0.17&  0.09&     4 &   2.07 &      --0.09&  0.12 &    5 &   2.15 \\
Zr II  &   0.05&   --  &    1 &   2.65 &       --0.44&   --  &    1 &   2.16&      --0.16&   -- &     1 &   2.44 &         -- &   --  &  --  &    --  \\
La II  &    -- &   --  &  --  &    --  &         0.05&   --  &    1 &   1.27&        0.06&   -- &     1 &   1.28 &        0.19&   --  &    1 &   1.41 \\
Ce II  & --0.22&  0.25 &    4 &   1.33 &         0.04&  0.20 &    4 &   1.59&      --0.17&  0.13&     2 &   1.38 &        0.08&  0.19 &    5 &   1.63 \\
Pr II  &    -- &   --  &  --  &    --  &       --0.20&   --  &    1 &   0.51&      --0.54&   -- &     1 &   0.17 &      --0.38&   --  &    1 &   0.33 \\
Nd II  &   0.06&  0.15 &    3 &   1.56 &       --0.14&  0.11 &    9 &   1.36&      --0.24&  0.12&     6 &   1.26 &      --0.06&  0.24 &    6 &   1.44 \\
Eu II  &   0.31&   --  &    1 &   0.82 &       --0.17&  0.00 &    2 &   0.34&      --0.27&  0.14&     2 &   0.24 &      --0.08&   --  &    1 &   0.43 \\
Gd II  &   --  &  --   & --   &   --   &         --  &  --   & --   &   --  &     --0.14 &  --  &    1  &  0.98  &        --  &  --   & --   &   --   \\
\hline
\end{tabular}
\begin{list}{}{}
\item[] N-- is the number of the line used in analysis.
\item[] [M/H] and (M/H) -- the relative-to-solar and absolute
abundances respectively.
\end{list}

{\bf Table A1 (continued)}\\
\tiny
\begin{tabular}{lrrrrrrrrrrrrrrrr}
\hline
\noalign{\smallskip}
\multicolumn{1}{c}{}&\multicolumn{4}{c}{EE Mon 4.80896}&
\multicolumn{4}{c}{XX Mon  5.45647}&\multicolumn{4}{c}{WW Pup   5.51672}&
\multicolumn{4}{c}{TZ Mon  7.42818}\\
\noalign{\smallskip}
\hline
 Ion &[M/H]
&$\sigma$&N&(M/H)&[M/H]&$\sigma$&N&(M/H)&[M/H]&$\sigma$&N&(M/H)&[M/H]&$\sigma$&N&(M/H)\\
\hline
Li I  &    -- &   --  &  --  &    --  &          --  &  -- &   --  &    --  &          -- &   --  &   -- &    -- &        --0.32: & -- &     1 &   0.84\\
C I   & --0.48&  0.13 &    5 &   8.07 &       --0.52 & 0.19&     2 &   8.03 &       --0.64&  0.16 &    5 &   7.91&        --0.53  &0.06&     3 &   8.02\\
N I   &   0.14&  0.05 &    2 &   8.11 &         0.15 & 0.09&     2 &   8.12 &       --0.16&   --  &    1 &   7.81&          0.26  & -- &     1 &   8.23\\
O I   &    -- &   --  &  --  &    --  &         0.06 & 0.04&     3 &   8.93 &          -- &   --  &   -- &    -- &          0.28  &0.08&     4 &   9.15\\
Na I  & --0.20&  0.18 &    4 &   6.13 &         0.02 & 0.04&     2 &   6.35 &         0.01&  0.11 &    2 &   6.34&          0.15  &0.09&     3 &   6.48\\
Mg I  & --0.60&   --  &    1 &   6.98 &          --  &  -- &   --  &    --  &       --0.33&   --  &    1 &   7.25&        --0.05  &0.13&     2 &   7.53\\
Al I  &    -- &   --  &  --  &    --  &         0.13 &  -- &     1 &   6.60 &       --0.12&  0.05 &    2 &   6.35&          0.05  &0.12&     5 &   6.52\\
Si I  & --0.36&  0.14 &   11 &   7.19 &       --0.18 & 0.05&     7 &   7.37 &       --0.13&  0.10 &   14 &   7.42&          0.01  &0.09&    25 &   7.56\\
Si II & --0.23&   --  &    1 &   7.32 &          --  &  -- &   --  &    --  &       --0.18&  0.08 &    2 &   7.38&        --0.11  & -- &     1 &   7.44\\
S I   & --0.56&  0.37 &    3 &   6.65 &       --0.21 & 0.24&     3 &  7.00  &       --0.20&  0.23 &    4 &   7.01&          0.20  &0.16&     4 &   7.41\\
Ca I  & --0.50&  0.15 &   12 &   5.86 &       --0.18 & 0.16&     6 &   6.18 &       --0.22&  0.07 &    4 &   6.14&          0.06  &0.15&     9 &   6.42\\
Sc I  &   0.07&   --  &    1 &   3.24 &         0.32 &  -- &     1 &   3.49 &         0.21&   --  &    1 &   3.38&        --0.23  &0.13&     4 &   2.94\\
Sc II & --0.50&  0.21 &    6 &   2.67 &       --0.31 & 0.06&     2 &   2.86 &       --0.48&  0.09 &    3 &   2.69&        --0.29  &0.22&     6 &   2.88\\
Ti  I & --0.17&  0.41 &    8 &   4.85 &       --0.22 & 0.10&     5 &   4.80 &       --0.24&  0.04 &    4 &   4.78&        --0.08  &0.11&    28 &   4.94\\
Ti II & --0.45&  0.20 &    4 &   4.57 &       --0.30 &  -- &     1 &   4.72 &       --0.30&   --  &    1 &   4.72&        --0.22  & -- &     1 &   4.80\\
V  I  &    -- &   --  &  --  &    --  &       --0.06 & 0.22&     4 &   3.94 &       --0.19&  0.15 &    4 &   3.81&        --0.08  &0.11&    19 &   3.92\\
V II  & --0.41&  0.32 &    2 &   3.59 &          --  &  -- &   --  &    --  &          -- &   --  &  --  &    -- &        --0.12  & -- &     1 &   3.88\\
Cr  I & --0.02&  0.64 &    6 &   5.65 &       --0.39 &  -- &     1 &   5.28 &       --0.14&  0.01 &    2 &   5.53&        --0.02  &0.15&    10 &   5.65\\
Cr II & --0.55&  0.09 &   10 &   5.12 &       --0.08 & 0.21&     5 &   5.59 &       --0.23&  0.16 &    5 &   5.44&        --0.18  &0.17&     2 &   5.49\\
Mn I  & --0.66&  0.19 &    2 &   4.73 &       --0.29 & 0.04&     4 &   5.10 &       --0.31&  0.08 &    2 &   5.08&          0.41  &0.07&     6 &   5.80\\
Fe I  & --0.51&  0.15 &   77 &   6.99 &       --0.18 & 0.11&    52 &   7.32 &       --0.17&  0.11 &   56 &   7.33&        --0.03  &0.15&   178 &   7.47\\
Fe II & --0.52&  0.17 &   29 &   6.98 &       --0.17 & 0.10&    15 &   7.33 &       --0.19&  0.12 &   19 &   7.31&        --0.04  &0.18&    22 &   7.46\\
Co I  &    -- &   --  &  --  &    --  &       --0.27 & 0.28&     2 &   4.65 &          -- &   --  &  --  &    -- &        --0.16  &0.06&    12 &   4.76\\
Ni I  & --0.58&  0.15 &   13 &   5.67 &       --0.27 & 0.08&    13 &   5.98 &       --0.21&  0.09 &   20 &   6.04&        --0.09  &0.10&    55 &   6.16\\
Cu I  &    -- &   --  &  --  &    --  &       --0.22 &  -- &     1 &   3.99 &       --0.34&  0.10 &    2 &   3.87&          0.17  & -- &     1 &   4.38\\
Zn I  &    -- &   --  &  --  &    --  &          --  &  -- &   --  &    --  &       --0.01&   --  &    1 &   4.59&          0.26  & -- &     1 &   4.86\\
Sr I  &    -- &   --  &  --  &    --  &          --  &  -- &   --  &    --  &          -- &   --  &  --  &    -- &        --0.10  & -- &     1 &   2.80\\
Y I   &    -- &   --  &  --  &    --  &          --  &  -- &   --  &    --  &         0.44&   --  &    1 &   2.68&        --0.47  & -- &     1 &   1.77\\
Y II  & --0.40&  0.20 &    4 &   1.84 &       --0.00 & 0.10&     5 &   2.24 &       --0.02&  0.09 &    6 &   2.22&        --0.08  &0.05&     2 &   2.16\\
Zr II &    -- &   --  &  --  &    --  &       --0.25 &  -- &     1 &   2.35 &          -- &   --  &  --  &    -- &        --0.22  & -- &     1 &   2.38\\
Ru I  &    -- &   --  &  --  &    --  &          --  &  -- &   --  &    --  &          -- &   --  &  --  &    -- &        --0.50  & -- &     1 &   1.34\\
Ba II & --0.51&   --  &    1 &   1.62 &          --  &  -- &   --  &    --  &          -- &   --  &  --  &    -- &          0.50  & -- &     1 &   2.63\\
La II &    -- &   --  &  --  &    --  &         0.06 &  -- &     1 &   1.28 &         0.15&   --  &    1 &   1.37&          0.04  & -- &     1 &   1.26\\
Ce II & --0.27&  0.51 &    2 &   1.28 &       --0.05 & 0.25&     4 &   1.50 &       --0.00&  0.16 &    3 &   1.55&        --0.26  &0.24&     3 &   1.29\\
Pr II &    -- &   --  &  --  &    --  &       --0.10 & 0.06&     2 &   0.61 &       --0.35&   --  &    1 &   0.36&           --   & -- &   --  &    -- \\
Nd II & --0.23&  0.30 &    2 &   1.27 &         0.01 & 0.24&     6 &   1.51 &       --0.04&  0.18 &    8 &   1.46&        --0.25  &0.02&     2 &   1.25\\
Eu II &    -- &   --  &  --  &    --  &         0.04 &  -- &     1 &   0.55 &       --0.09&  0.05 &    2 &   0.42&        --0.01  & -- &     1 &   0.50\\
\hline
\hline
\multicolumn{1}{c}{}&\multicolumn{4}{c}{AC Mon  8.01425}&
\multicolumn{4}{c}{TX Mon  8.70173}&\multicolumn{4}{c}{HW Pup  13.454}&
\multicolumn{4}{c}{AD Pup  13.5940}\\
\noalign{\smallskip}
\hline
 Ion &[M/H]
&$\sigma$&N&(M/H)&[M/H]&$\sigma$&N&(M/H)&[M/H]&$\sigma$&N&(M/H)&[M/H]&$\sigma$&N&(M/H)\\
\hline
Li I   &    --  &  --  &   -- &    --  &   -- &   -- &    --&     -- &          -- &   -- &   -- &     -- &         0.18:&  -- &     1&    1.34\\
C I    & --0.52 & 0.15 &    7 &   8.03 &--0.41&  0.07&    10&    8.14&       --0.56&  0.11&     5&    7.99&       --0.77 & 0.00&     2&    7.78\\
N I    &   0.23 & 0.14 &    5 &   8.20 &  0.29&  0.10&     6&    8.26&         0.04&  0.12&     3&    8.01&         0.49 &  -- &     1&    8.45\\
O I    & --0.25 &  --  &    1 &   8.62 &--0.16&  0.31&     2&    8.71&       --0.29&   -- &     1&    8.58&          --  &  -- &   -- &     -- \\
Na I   &   0.05 & 0.18 &    2 &   6.38 &--0.08&  0.10&     3&    6.25&       --0.20&   -- &     1&    6.13&         0.06 & 0.18&     2&    6.39\\
Mg I   &    --  &  --  &   -- &    --  &--0.26&  0.20&     2&    7.32&       --0.52&   -- &     1&    7.06&          --  &  -- &   -- &     -- \\
Al I   & --0.36 & 0.15 &    2 &   6.11 &--0.07&  0.21&     4&    6.40&       --0.16&   -- &     1&    6.31&         0.00 &  -- &     1&    6.47\\
Si I   & --0.08 & 0.12 &   10 &   7.47 &--0.10&  0.08&    27&    7.45&       --0.20&  0.08&    11&    7.35&       --0.19 & 0.10&     9&    7.36\\
Si II  &    --  &  --  &   -- &    --  &  0.06&   -- &     1&    7.61&       --0.07&   -- &     1&    7.49&       --0.21 &  -- &     1&    7.34\\
S I    & --0.21 & 0.19 &    6 &  7.00  &--0.17&  0.25&     4&    7.04&       --0.29&  0.30&     3&    6.92&       --0.13 & 0.23&     2&    7.08\\
Ca I   & --0.13 & 0.15 &    7 &   6.23 &--0.24&  0.04&     6&    6.12&       --0.27&  0.15&     6&    6.09&       --0.17 & 0.16&     3&    6.19\\
Sc I   &    --  &  --  &   -- &    --  &   -- &   -- &    --&     -- &          -- &   -- &    --&     -- &       --0.04 & 0.15&     2&    3.13\\
Sc II  & --0.21 & 0.19 &    3 &   2.96 &--0.16&  0.13&     3&    3.01&       --0.30&  0.13&     3&    2.87&          --  &  -- &   -- &     -- \\
Ti  I  & --0.28 &  --  &    1 &   4.74 &--0.16&  0.06&     3&    4.86&       --0.22&  0.16&     2&    4.80&       --0.20 & 0.19&     5&    4.82\\
Ti II  & --0.15 & 0.09 &    2 &   4.87 &--0.19&  0.09&     2&    4.83&       --0.30&   -- &     1&    4.72&       --0.31 &  -- &     1&    4.71\\
V  I   & --0.17 &  --  &    1 &   3.83 &--0.28&  0.12&     2&    3.72&       --0.09&   -- &     1&    3.91&       --0.32 & 0.16&     6&    3.68\\
V II   & --0.29 &  --  &    1 &   3.71 &--0.27&  0.04&     2&    3.73&       --0.45&  0.16&     2&    3.55&       --0.35 &  -- &     1&    3.65\\
Cr  I  & --0.54 &  --  &    1 &   5.13 &--0.17&  0.06&     5&    5.50&       --0.45&   -- &     1&    5.22&       --0.54 &  -- &     1&    5.13\\
Cr II  & --0.20 & 0.07 &    7 &   5.47 &--0.15&  0.09&     3&    5.52&       --0.27&  0.14&     7&    5.40&       --0.30 & 0.20&     2&    5.37\\
Mn I   & --0.25 & 0.14 &    5 &   5.14 &--0.25&  0.06&     7&    5.14&       --0.28&  0.08&     3&    5.11&       --0.33 & 0.21&     3&    5.06\\
Fe I   & --0.22 & 0.07 &   57 &   7.28 &--0.14&  0.06&   118&    7.36&       --0.30&  0.10&    53&    7.20&       --0.24 & 0.08&    46&    7.26\\
Fe II  & --0.20 & 0.07 &   15 &   7.30 &--0.14&  0.07&    23&    7.36&       --0.28&  0.15&    27&    7.22&       --0.21 & 0.09&     6&    7.29\\
Co I   &    --  &  --  &  --  &    --  &--0.24&  0.08&     3&    4.68&       --0.18&   -- &     1&    4.74&       --0.48 & 0.06&     3&    4.44\\
Ni I   & --0.35 & 0.06 &    4 &   5.90 &--0.20&  0.07&    30&    6.05&       --0.35&  0.10&    10&    5.90&       --0.37 & 0.07&    23&    5.88\\
Cu I   & --0.80 &  --  &    1 &   3.41 &--0.04&  0.05&     2&    4.17&       --0.32&  0.14&     2&    3.89&       --0.47 &  -- &     1&    3.74\\
Zn I   &    --  &  --  &  --  &    --  &  0.15&   -- &     1&    4.75&          -- &   -- &   -- &     -- &          --  &  -- &   -- &     -- \\
Y II   & --0.12 & 0.13 &    5 &   2.12 &  0.06&  0.10&     5&    2.30&       --0.13&  0.07&     6&    2.11&       --0.11 & 0.05&     3&    2.13\\
Zr II  &    --  &  --  &  --  &    --  &--0.27&   -- &     1&    2.33&          -- &   -- &   -- &     -- &       --0.20 &  -- &     1&    2.40\\
La II  &   0.13 &  --  &    1 &   1.35 &  0.23&   -- &     1&    1.45&          -- &   -- &   -- &     -- &         0.27 &  -- &     1&    1.49\\
Ce II  & --0.20 & 0.18 &    2 &   1.35 &  0.10&  0.15&     3&    1.65&       --0.13&  0.24&     5&    1.42&       --0.14 & 0.22&     3&    1.41\\
Pr II  &    --  &  --  &  --  &    --  &--0.12&   -- &     1&    0.59&       --0.28&  0.07&     3&    0.43&       --0.29 &  -- &     1&    0.42\\
Nd II  & --0.12 & 0.39 &    2 &   1.38 &--0.10&  0.13&     5&    1.40&       --0.15&  0.22&     8&    1.35&       --0.12 & 0.22&     6&    1.38\\
Eu II  &   0.18 & 0.11 &    2 &   0.69 &  0.07&   -- &     1&    0.58&       --0.11&  0.12&     2&    0.40&       --0.06 &  -- &     1&    0.45\\
Gd II  &    --  &  --  &  --  &    --  &--0.23&   -- &     1&    0.89&          -- &   -- &    --&     -- &       --0.03 &  -- &     1&    1.09\\
\hline
\end{tabular}

{\bf Table A1 (continued)}\\
\tiny
\begin{tabular}{lrrrrrrrrrrrr}
\hline
\hline
\noalign{\smallskip}
\multicolumn{1}{c}{}&\multicolumn{4}{c}{BN Pup  13.6731}&
\multicolumn{4}{c}{SV Mon 15.23278}&\multicolumn{4}{c}{VZ Pup  23.1710}\\
\noalign{\smallskip}
\hline
 Ion &[M/H]
&$\sigma$&N&(M/H)&[M/H]&$\sigma$&N&(M/H)&[M/H]&$\sigma$&N&(M/H)\\
\hline
C I    & --0.45&  0.08&     2&8.10& --0.81&   -- &  1&    7.74&--0.83&  0.18&     5&    7.72 \\
N I    &   0.55&  0.08&     2&8.52&$<$0.22&   -- &  1& $<$8.19&  0.02&  0.16&     7&    7.99 \\
O I    & --0.06&  0.04&     3&8.81&   0.21&  0.07&  3&    9.08&--0.10&   -- &     1&    8.77 \\
Na I   &   0.09&  0.02&     2&6.42&   0.10&   -- &  1&    6.43&  0.00&  0.08&     2&    6.33 \\
Mg I   &   0.21&  0.05&     3&7.79& --0.48&   -- &  1&    7.10&--0.15&  0.15&     4&    7.43 \\
Al I   &   0.04&  0.11&     4&6.51& --0.04&  0.12&  6&    6.43&--0.15&  0.11&     2&    6.32 \\
Si I   & --0.01&  0.11&    21&7.54& --0.02&  0.13& 23&    7.53&--0.11&  0.09&    13&    7.44 \\
Si II  &   0.08&   -- &     1&7.63&   0.05&   -- &  1&    7.60&  0.20&   -- &     1&    7.75 \\
S I    &   0.19&  0.16&     4&7.40&   0.24&  0.24&  5&    7.45&--0.22&  0.21&     3&    6.99 \\
Ca I   &   0.04&  0.24&     6&6.40& --0.13&  0.07&  2&    6.23&--0.12&  0.09&     7&    6.24 \\
Sc I   &   0.05&   -- &     1&3.22& --0.07&  0.09&  2&    3.10&   -- &   -- &   -- &     --  \\
Sc II  & --0.08&   -- &     1&3.09& --0.07&   -- &  1&    3.10&--0.16&  0.20&     3&    3.01 \\
Ti  I  & --0.05&  0.13&    13&4.97& --0.11&  0.17& 17&    4.91&--0.23&  0.00&     2&    4.79 \\
Ti II  & --0.02&   -- &     1&5.00&   0.11&   -- &  1&    5.13&--0.33&   -- &     1&    4.69 \\
V  I   & --0.12&  0.10&    11&3.88& --0.22&  0.08& 12&    3.78&  0.12&  0.01&     2&    4.12 \\
V II   & --0.24&  0.06&     3&3.76& --0.09&  0.12&  3&    3.91&--0.26&   -- &     1&    3.74 \\
Cr  I  &   0.07&  0.26&     4&5.74& --0.06&  0.21&  6&    5.61&--0.45&   -- &     1&    5.22 \\
Cr II  &    -- &   -- &    --& -- & --0.02&   -- &  1&    5.65&--0.12&  0.09&     5&    5.55 \\
Mn I   & --0.02&  0.05&     6&5.37&   0.03&  0.15&  4&    5.42&--0.28&  0.04&     5&    5.11 \\
Fe I   &   0.01&  0.06&    85&7.51&   0.00&  0.06&115&    7.50&--0.16&  0.09&    58&    7.34 \\
Fe II  &   0.01&  0.13&    19&7.51&   0.02&  0.04& 12&    7.52&--0.15&  0.08&    15&    7.35 \\
Co I   & --0.23&  0.08&     8&4.69& --0.24&  0.11&  8&    4.68&   -- &   -- &   -- &     --  \\
Ni I   & --0.03&  0.09&    33&6.22& --0.08&  0.07& 33&    6.17&--0.28&  0.08&    10&    5.97 \\
Cu I   & --0.24&   -- &     1&3.97& --0.42&   -- &  1&    3.79&  0.10&   -- &     1&    4.31 \\
Y I    &   0.50&   -- &     1&2.74&   0.19&   -- &  1&    2.43&   -- &   -- &   -- &     --  \\
Y II   &   0.15&  0.05&     3&2.39&   0.16&  0.04&  2&    2.40&--0.06&  0.06&     4&    2.18 \\
Zr II  & --0.17&   -- &     1&2.43&   0.00&   -- &  1&    2.60&   -- &   -- &   -- &     --  \\
La II  &   0.30&   -- &     1&1.52&   0.30&   -- &  1&    1.52&   -- &   -- &   -- &     --  \\
Ce II  & --0.00&  0.14&     4&1.55&   0.08&  0.20&  3&    1.63&--0.11&  0.28&     3&    1.44 \\
Pr II  & --0.15&  0.02&     2&0.56&    -- &   -- &-- &     -- &--0.21&   -- &     1&    0.50 \\
Nd II  &   0.03&  0.18&     8&1.53&   0.16&  0.25&  3&    1.66&  0.02&  0.22&     4&    1.52 \\
Eu II  &   0.07&  0.13&     2&0.58&   0.08&  0.07&  2&    0.59&  0.02&   -- &     1&    0.53 \\
Gd II  &    -- &   -- &    --& -- &   0.12&   -- &  1&    1.24&  0.32&   -- &     1&    1.44 \\
\hline
\end{tabular}
\end{appendix}


\begin{thebibliography}{}
\bibitem[1997]{amle97}
 Amaral L.H., L\'epine J.R.D., 1997, MNRAS 286, 885
\bibitem[2002]{andret02a}
 Andrievsky S.M., Kovtyukh V.V., Luck R.E., L\'epine J.R.D., Bersier D.,
  Maciel W.J., Barbuy B., Klochkova V.G., Panchuk V.E., Karpischek R.U.,
  2002a, A\&A 381, 32
\bibitem[2002]{andret02b}
 Andrievsky S.M., Bersier D., Kovtyukh V.V., Luck R.E., Maciel W.J.,
  L\'epine J.R.D., Beletsky Yu.V., 2002b, A\&A 384, 140
\bibitem[2001]{capet01}
 Caputo F., Marconi M., Musella I., Pont F., 2001, A\&A 372, 544
\bibitem[1995] {chdp95}
 Christensen-Dalsgaard , Petersen J.O., 1995, A\&A 299, L17
\bibitem[1995]{feret95}
Fernie J.D., Evans N.R., Beattie B., Seager S., 1995, IBVS 4148, 1
\bibitem[1995]{fri95}
 Friel E.D., 1995, ARA\&A 33, 381
\bibitem[1998]{gfg98}
Gieren W.P., Fouqu\'e P., G\'omez M., 1998, ApJ 496, 17
\bibitem[1979]{jan79}
 Janes K.A., 1979, ApJS 39, 135
\bibitem[1993]{ls93}
Laney C.D., Stobie R.S., 1993, MNRAS 263, 291
\bibitem[2001]{lmd01}
 L\'epine J.R.D., Mishurov Yu.N., Dedikov S.Yu., 2001, ApJ 546, 234
\bibitem[1977]{luck77}
 Luck  R.E. 1977, ApJ 218, 752
\bibitem[2002]{mishet02}
 Mishurov Yu.N., L\'epine J.R.D., Acharova I.A. 2002, astro-ph/0203458
\bibitem[1981]{pantos81}
 Panagia N., Tosi M., 1981, A\&A 96, 306
\bibitem[1999a]{sev99a}
 Sevenster M.N. 1999a, MNRAS 310, 629
\bibitem[1999b]{sev99b}
 Sevenster M.N. 1999b, ApSS 265, 377
\bibitem[1997]{twaet97}
 Twarog B.A., Ashman K.M., Antony-Twarog B.J., 1997, AJ 114, 2556
\end{thebibliography}
\end{document}